# Effects of hydrogen bonding on supercooled liquid dynamics and the implications for supercooled water

J. Mattsson[*], R. Bergman, P. Jacobsson & L. Börjesson

Department of Applied Physics, Chalmers University of Technology, SE-412 96, Göteborg, Sweden

[*]Corresponding author: johanm@chalmers.se

**Abstract**

The supercooled state of bulk water is largely hidden by unavoidable crystallization, which creates an experimentally inaccessible temperature regime - a 'no man's land'. We address this and circumvent the crystallization problem by systematically studying the supercooled dynamics of hydrogen bonded oligomeric liquids (glycols), where water corresponds to the chain-ends alone. This novel approach permits a 'dilution of water' by altering the hydrogen bond concentration via variations in chain length. We observe a dynamic crossover in the temperature dependence of the structural relaxation time for all glycols, consistent with the common behavior of most supercooled liquids. We find that the crossover becomes more pronounced for increasing hydrogen bond concentrations, which leads to the prediction of a marked dynamic transition for water within 'no man's land' at $T \sim 220$ K. Interestingly, the predicted transition thus takes place at a temperature where a so called 'strong-fragile' transition has previously been suggested. Our results, however, imply that the dynamic transition of supercooled water is analogous to that commonly observed in supercooled liquids. Moreover, we find support also for the existence of a secondary relaxation of water with behavior analogous to that of the secondary relaxation observed for the glycols.

Water like most liquids can be supercooled, which means that it can be cooled below its freezing point without crystal formation. The hallmark of a supercooled liquid is a dramatic slowing down of molecular dynamics for decreasing temperatures, and when reaching the glass transition temperature, $T_g$, the sample falls out of equilibrium on experimental time-scales. The result is a disordered solid - a glass. Water can be supercooled to $T\sim235$ K, where it inevitably crystallizes due to homogeneous nucleation [1-2]. A small quantity of water can, however, be rapidly quenched to temperatures below 100 K into its glassy state. Upon heating the glass, unavoidable crystallization sets in at $T\sim150$ K. Consequently, a large experimentally inaccessible temperature regime exists for bulk liquid water, stretching from ~150 K to 235 K. Due to this 'no man's land' many of water's properties are still unexplained.

It is unclear at what temperature supercooled water forms a glass. A glass-liquid transition at $124\leq T_g\leq136$ K [3-4] has often been reported, but this result has recently been debated [5-7] and a second temperature range around 165±5 K, thus within 'no man's land', has also been suggested. Furthermore, it is well established that the temperature dependence of the viscosity, and the accompanying structural relaxation time, follows a power law-like behavior at high temperatures with an apparent divergence at a temperature $T_s\approx215$-230 K, hidden within the inaccessible temperature range [*1-2*]. Since $T_s$ is clearly different from any estimate of $T_g$, a change of behavior must take place within 'no man's land' and the existence of a dynamic crossover, a so called 'strong-fragile' transition [*8*], has been proposed. Several explanations for these observations have been suggested [*1-2*], including the existence of a second thermodynamic critical point, which if it exists could make water highly exclusive among liquids.

The existence of a dynamic crossover is not, however, restricted to water. Supercooled liquids generally show different dynamic behavior above and below a crossover temperature $T^* \approx 1.1\text{-}1.6 \cdot T_g$, This crossover is observed in a range of dynamical properties. For instance, the temperature dependence of the structural relaxation time changes and a decoupling takes place between translational and rotational motion [*9-10*]. For $T<T^*$, the viscosity or the related structural α-relaxation time, $\tau_\alpha$, show different sensitivity to a temperature change for different liquids; this is captured by the notion of fragility, which is often quantified as the slope in a $T_g$-scaled Arrhenius plot at $T_g$, $m=d\log(\tau_\alpha)/d(T_g/T)\,|_{T=Tg}$. Liquids with high values of *m*, often showing markedly non-Arrhenius temperature dependence of their structural relaxation time, $\tau_\alpha$, are called *fragile*, wheareas liquids with small values of *m*, where $\tau_\alpha$ follow a behavior close to the Arrhenius law ($\ln(\tau) \propto 1/T$) are termed *strong*. Fragile liquids are found among weakly interacting ionic and van der Waals systems, while strong liquids are typically network glass-formers such as $SiO_2$ or $GeO_2$. The temperature variation of the structural relaxation time is generally well described in this dynamic regime by the empirical Vogel-Fulcher-Tammann (VFT) equation [*9*], $\tau_\alpha=\tau_0\exp(DT_0/(T-T_0))$, where $T_0$ is a divergence temperature, *D* describes the increase of $\tau_\alpha$ on approach of $T=T_0$, and $\tau_0$ is the relaxation time in the high-T limit. For $T>T^*$, supercooled liquids generally show stronger deviations from Arrhenius behavior than for $T<T^*$ [10] and the temperature dependence of the structural relaxation time near, but above, $T^*$ is often described with a power law expression, $\tau_\alpha \propto (T-T_c)^{-\gamma}$ [9], where the temperature of divergence, $T_c \sim T^*$. This is reminiscent of the behavior observed for supercooled water [11]. Mode coupling theory has often been invoked to explain this type of behavior in supercooled liquids [9,11]. These observations, together with the fact that many properties of water result mainly from its high density of H-bonds, emphasize the importance of investigating the role of H-bonding for the supercooled dynamics in glass-forming liquids in general.

We here study a series of glass-forming supercooled liquids, for which the hydrogen bond density can be systematically varied and where water is a member. We investigate homologous oligomers with the same monomeric unit, propylene oxide, but varying in chain-ends: X-[CH$_2$CH(CH$_3$)O]$_N$-Y, where {X=HO;Y=H} for glycols, {X=CH$_3$-O;Y=H} for monomethyl ethers (MMEs) [12], and {X=CH$_3$-O; Y=CH$_3$} for dimethyl ethers (DMEs), respectively and N denotes the number of monomer units in the chain. All samples were carefully dried before the measurements to avoid any effects due to water contamination. The molecules formed solely by the end-groups, thus corresponding to N=0, are water, methanol and dimethyl ether for the glycols, MME:s and DME:s, respectively. H-bonding plays an important role for glycols and MMEs, while the DMEs lack the possibility of such interactions, thus providing the perfect non-H-bonded reference system for this study.

We follow the molecular dynamics over a broad frequency ($\sim$10$^{-2}$-10$^9$ Hz) and temperature range using a high resolution dielectric spectrometer (Novocontrol). Data for the glycol trimer is shown near $T_g$ in Fig. 1a, clearly demonstrating the presence of a primary structural α- and a secondary β-relaxation. Asymmetric (KWW-type) and symmetric (Cole-Cole) response functions [13] were fitted to the two relaxation peaks respectively and the molecular relaxation times were determined as $\tau_\alpha = 1/(2\pi f_p)$, where $f_p$ denotes the peak frequencies. The relaxation times as determined for the trimer is shown in Fig. 1b showing the non-Arrhenius α- and the Arrhenius β-relaxation.

The α-relaxation times near $T_g$ are shown for the three polyether series in Fig. 1c-e. The temperature dependence of the $T_g$-scaled relaxation times are nearly chain-length independent for the $CH_3$-terminated chain series (DMEs), showing that no significant intermolecular interactions act via the chain-ends. In contrast, the presence of hydroxyl end-groups and thus hydrogen bonding for the MMEs, with OH-termination at one chain-end, and glycols, with OH-termination at both chain-ends, markedly changes the temperature behavior of the structural relaxation. For both chain-series, a decreasing chain-length, and thus increased hydroxyl group density, leads to a systematic approach towards an Arrhenius behavior. These results clearly stress the importance of H-bonds for the relaxation dynamics near $T_g$.

In Fig. 2a, we show the α-relaxation times for the glycols over a wide range of temperatures and time-scales. The pronounced chain-length dependence of the dynamics at low temperatures is clearly observed, but the chain-length dependence at higher temperatures is much weaker. To better investigate this behavior, we plot the parameter $Z=100 \cdot [-d\log(f_p)/d(1/T)]^{-1/2}$ vs inverse temperature in Fig. 2b, where $f_p=1/(2\pi\tau_\alpha)$ is the α peak frequency from dielectric spectroscopy. This representation linearizes the VFT behavior [*10*], which describes the data well at low temperatures.

For all glycols, we find two separate dynamic regimes, I & II, with a crossover at a temperature, $T^*$ [*14*]. At low $T$, in regime II, all glycols are well described by straight lines and thus VFT functions, and the approach towards an Arrhenius behavior for the shortest glycols is clearly observed as a decrease in the negative slope of the linearized VFTs; an Arrhenius behavior corresponds to a horizontal line. The magnitude of the negative slope in this kind of plot is thus a good quantification of the deviation from Arrhenius behavior [*15*]. At high temperatures above $T^*$, in regime I, the glycols show a highly non-Arrhenius behavior, which is not markedly affected by a variation in H-group density over the investigated temperature range. Thus, for all chain-lengths we find a clearly non-Arrhenius relaxation behavior at high $T$, while shorter chains approach an Arrhenius behavior for $T<T^*$. The subtle change of $T$-dependence observed for long chain glycols at $T^*$ turns into a significant crossover for sufficiently short chains.

Since water corresponds to the glycol with $N=0$, we include dielectric results for water at high $T$ in Fig. 2a and b, and we see that its highly non-Arrhenius power law-like behavior is similar to the distinctly non-Arrhenius behavior of the glycols in regime I. Due to crystallization, we can not measure bulk water in regime II. However, as shown in Fig. 2b all glycol data in regime II converge at $T^*$ and are well described by $Z^*-Z=S(T^*/T-1)$, with $Z^*=1.89\pm0.04$. The variation of the negative slope $S$ with molecular weight $M$, and thus H-bond density, is highly systematic as demonstrated in Fig. 2c. We extrapolate to find $S=0.65$ for the molecular weight of water. We thus predict that the structural relaxation time of water is significantly more Arrhenius-like, than the longer chain glycols for $T<T^*$. Thus, water should undergo a marked dynamic transition within 'no man's land', in accordance with a suggested transition from highly fragile to less fragile (stronger) behavior [8]. The existence of a dynamic transition is, however, not exclusive to water but it occurs also in other related H-bonded liquids, as observed in Fig. 2. In addition, similar results are found for many other supercooled liquids [10], which suggests that the phenomenon is of a general character, even if the transition for water should be unusually well pronounced if it could be directly measured.

Interestingly, as shown in Fig. 2c, the $T_g$-normalized crossover temperature, $T^*/T_g$, varies systematically with chain-length. Since $T^*$ for water should lie within the range 216-235 K [17], the two postulated temperature ranges for $T_g$, discussed above, gives two possible ranges for the $T^*/T_g$ ratio for water, as shown in Fig. 2c. Even though a direct extrapolation of the ratio for water is speculative, it is clear that our data is consistent with the lower of the two temperature ranges $124 \leq T_g \leq 136$ K [3-4]. Further support for these conclusions are obtained from an investigation of the $T_g$-values for the three chain-length series, as shown in Fig. 3. The DME:s show a smooth monotonic decrease of $T_g$ for shorter chains, as expected for an oligomer-system without strong intermolecular interactions. The MMEs show a very similar dependence, however shifted to shorter chain lengths. This is readily understood since we expect the major interaction in OH-capped polyethers to be due to H-bonds between the end-groups [18]. MME:s with length N thus pair up and correspond to a DME of length $\sim 2N$. Rescaling the abscissa by a factor 2 for the MME $T_g$:s indeed leads to good agreement with those of the DMEs, as shown in the inset to Fig. 3. The $T_g$-behavior for the glycols is also consistent with a formation of effective chains.

Rescaling the abscissa by a constant factor (=8) shifts the data onto the mastercurve formed by the DMEs and the rescaled MMEs. We note, however, that an additional small oscillation is present. The result suggests that effective chains are formed by joining oligomers via, on average, seven end-end-links. Since our analysis supports $124 \leq T_g \leq 136$ K, we add water in the plot as the "$N$=0 glycol". We find a good correspondance with the behavior of the polyethers. Intriguingly, this suggests a correlation between the H-bond coordination of water and the glycols in the dynamic regime near the glass transition. Furthermore, we use the extrapolated $S$=0.65, a $T_g$=133 K and $T^*$=216 K [19] to estimate the VFT behavior for water in regime II. The result, displayed as the solid line in Fig. 4, corresponds to an $m$-value of ~34, typical of a moderately fragile liquid. Thus our scaling prediction is that supercooled water undergoes a marked dynamic crossover at $T^*$, from a highly non-Arrhenius to a near Arrhenius behavior. If the high temperature behavior had persisted until the glass-transition was reached, it would correspond to a highly fragile liquid (very high $m$). Instead, however, at the crossover a change into a stronger, but still moderately fragile liquid takes place. Our result is consistent with dielectric relaxation data on water sequestered in a hydrogel [4], as indicated by the open circles in Fig. 4. Recent self-diffusion data [21], obtained from isotope intermixing experiments [22-23], also suggest a very similar picture. We note, however, that the interpretations of the latter experiments have recently been questioned [24].

Glass formers generally display a secondary so called $β$-relaxation [9] in addition to the structural α-relaxation. If water behaves as other supercooled liquids, we would expect it to show also this relaxation. Fig. 5a shows the $β$-relaxations for the different oligomer series. In previous studies of the $β$-relaxation in DMEs [25] we have shown that the relaxation is of a cooperative nature and involves 4-5 monomer units along the oligomer backbone. Shorter chains result in a speed-up of the dynamics, as seen in Fig. 5a and b. The MME:s display a highly similar behavior, which is expected, due to H-bonding between end-groups, as discussed above. The glycols, however, behave qualitatively different, with systematically slower $\beta$-relaxations for shorter chains. A full account of these results will be presented in a forthcoming article, but as briefly outlined in [*26*], the data is fully consistent with the results for the $\alpha$-relaxation, and suggests end-end linked glycol chains with an effective length > 4-5 monomer units. Focusing on the glycol $\beta$-relaxations, we find that the relaxation times increase regularly with decreasing chain-length, or correspondingly increasing H-bond concentration, as shown in Fig. 5b. This observation implies the existence of a $\beta$-relaxation of the same type for water, with a relaxation time consistent with the extrapolation shown in Fig. 5b.

Remarkably, support of this prediction is found experimentally. In Fig. 5a we have in addition to the glycol $\beta$-relaxations included the relaxation measured for water confined in a Na-vermiculate clay [*27*], where crystallization is avoided. We note that the β-relaxation dynamics of the glycols has been found to be essentially undisturbed by this confinement [*28*]. The correspondence between the prediction and the data is striking, suggesting that the relaxation found in confined water is a $\beta$-relaxation of the same nature as that measured in the polyethers [*29*]. In addition, it has recently been demonstrated that a secondary relaxation with properties nearly identical to the data in Fig. 5a is found for water subjected to a large range of different types of confinements, where measurements are possible in the deeply supercooled state due to suppressed crystallization [*30*]. This further supports that the relaxation is not strongly sensitive to the specific confinement, but is most likely a generic feature of water itself. In analogy to glass-forming liquids in general [9], our study thus suggests a relaxation scenario for water including both a structural α- and a secondary β-relaxation as shown in Fig. 4. Since the separation between the two predicted relaxations is small near $T_g$, however, we expect the secondary process to be observed only as an "excess wing" on the high-frequency flank of the α-relaxation [31], analogous to the behavior observed for the glycol monomer (see caption to Fig. 5). Dielectric measurements in the glassy state of water has indeed given evidence for the presence of a secondary process, largely submerged by the α-relaxation [32].

To explain the behavior of supercooled water, a range of models rather specific to water have been invoked [2]. However, the strong similarities presented in this work between supercooled dynamics in water and other hydrogen bonded liquids and even between water and liquids in general, either challenge these models or conversely suggest that the models might be applicable to a much larger family of liquids. It is for instance relevant to ask if the unusual behavior observed in the thermodynamic response functions of water [2] might be reflected also in the response of other hydrogen bonded liquids. Some indications suggest that this might indeed be the case [33-34].

To fully resolve these questions, it is essential to understand hydrogen bonding itself and specifically its effect on liquids in the supercooled state. Hydrogen bonding is still a highly debated topic even for water in its normal liquid state. The traditional picture for liquid water [1-2] suggests a tetrahedral bonding structure on average. This view was recently challenged [35] and in this debated [36] interpretation it is instead suggested that most molecules are coordinated through one 'donor' and one 'acceptor' hydrogen bond. Our finding of a link between the supercooled dynamics of water and polyethers further highlights the need for more detailed investigations of hydrogen bond coordinations and their role in supercooled dynamics, not only in water, but in hydrogen bonded liquids in general.

To conclude, we have shown that the dynamic crossover for supercooled oligomeric glycols becomes systematically more pronounced for higher H-bond concentrations. This leads to a prediction of a significant dynamic crossover for water within 'no man's land' at T~220 K, in the same temperature range where a so called 'strong-fragile' transition has previously been suggested. Our work thus implies that the dynamic transition of supercooled water has the same origin as the dynamic crossover commonly observed in supercooled liquids. In addition, we find support for the existence of a secondary relaxation in water, with a behavior analogous to that of the secondary relaxation observed for the glycols.

**Acknowledgements**

The authors are grateful for financial support from the Knut and Alice Wallenberg Foundation and from the Swedish Research Council. We thank Maurizio Furlani and Daniel Pettersen for help with sample synthesis and are greatful to the group of Prof. Juan Colmenero for assistance with some complementary dielectric measurements.

[*13*] J. Mattsson, R. Bergman, P. Jacobsson and L. Börjesson, *Phys. Rev. Lett.* **90**, 075702-1 (2003).

[*14*] As discussed in the text, the data at low temperatures can be well described by a VFT expression and thus a straight line in the linearized representation of Fig. 2b. To estimate a characteristic temperature, T*, for the crossover we approximate the high temperature regime with a straight line, and determine the crossover temperature as the intersection of the two lines. We estimate that we can generally determine $T^*$ within ±2-3 K, as shown from the error-bars in Fig. 2c ($T_g$ is determined within ±1 K). We emphasize that we do not assign any physical significance to this choice of fitting function for the high temperature regime, but only use it as a straightforward means to consistently extract the crossover temperature. For the glycol tetramer, we have data only near $T_g$. However, since the crossover temperature is $T^*$=282±3 K for the longer chain glycols, $T^*$=282 K is used.

[*15*] The standard way of parameterizing the VFT behavior as $\tau_\alpha = \tau_0 \cdot \exp[D \cdot T_0/(T-T_0)]$ is often problematic; $\tau_0$ is the time-scale at infinite temperatures and is thus defined outside the range of validity of the VFT expression and at $T_0$ the system is out of equilibrium. This parameterization thus yields parameters without a clear physical validity, which for hydrogen bonded systems often becomes evident, for instance reflected in $\tau_0$ values that are inconsistent with a molecular vibrational time-scale. In order to circumvent this problem, we here choose an alternative parameterization of the VFT-expression. Instead of the parameters [$\tau_0,D,T_0$], given $T^*$ we can alternatively use [$\tau^*,Z^*,S$], where $\tau^*$ is the time-scale at the crossover temperature, $Z^*$ through its

definition is directly related to the slope in an Arrhenius plot at $T^*$ and $S$ is a direct measure of the deviation from an Arrhenius behavior. This parameterization also has the direct advantage that the parameters are defined within the region of validity of the VFT expression. Interestingly, $\tau^*$ is generally ~$10^{-8}$-$10^{-7}$ s for supercooled liquids [*16*].

[*16*] V.N. Novikov and A.P. Sokolov, *Phys. Rev. E* **67**, 031507 (2003).

[*17*] A lower limit for the crossover temperature, $T^*$, is determined from the power law divergence near 216 K, estimated from a fit to dielectric data at high temperatures, as shown in Fig. 4. An upper limit is set by the homogeneous nucleation temperature at ~ 235 K, which measured properties such as the thermodynamic response functions and the structural relaxation time, approach in a power law-like manner.

[*18*] A. Bernson and J. Lindgren, *Polymer* **35**, 4842 (1994).

[*19*] This value is consistent both with the behavior of the $T^*/T_g$ ratio in Fig. 2c and with the fact that the structural relaxation time at $T^*$ varies only slightly between the glycol samples (compare with results in [20]) and for water it extrapolates to ~$6 \cdot 10^{-9}$ s.

[*20*] A. Schönhals, *Europhys. Lett.* **56**, 815 (2001).

[*21*] The data set and our scaling estimate provide consistent pictures, both with

regards to $T_g$-values and apparent activation energies [22-23].

[*29*] The dielectric loss is well described by a Cole-Cole expression, which is typical for secondary relaxations [*25*] and gives further evidence for an assignment of the dielectric relaxation as a *β*-relaxation.

[*39*] The $T_g$ of dimethyl ether was estimated from the empirical finding that the ratio of the boiling and glass transition temperatures, $T_b/T_g$~3.3 for the samples where it could be determined. Given the known $T_b$ (248 K) of dimethyl ether, $T_g$ was estimated to 76 K.

[*40*] P. Ayotte, R. Scott Smith, G. Teeter, Z. Dohnálek, G.A. Kimmel *et al.*, *Phys. Rev. Lett.* **88**, 245505-1 (2002).

**Figure captions**

**Figure 1 | Supercooled relaxation dynamics for polyethers. a**. Dielectric permittivity $\varepsilon''$ vs frequency for temperatures from 134 K to 240 K, for the glycol trimer. The structural α- and the secondary β-relaxation are visible as clear loss peaks. **b.** The relaxation times for the α- and β-relaxations determined from the data in a, as described in the text. **c-e**. α-relaxation times versus inverse temperature normalized to $T_g$ for the three investigated polyether series at temperatures near $T_g$. The symbols represent the number of monomer units, *N*, for each sample in a series with N=1 (○) , N=2 (□), N=3 (△), N=4 (◇), N=7 (▷) and N=70 (▽). The N=70 glycol data have been included in all panels, c-e, to facilitate comparisons. The solid lines are VFT fits to the data. *R* in the chemical formulae represents a -$CH_2CH(CH_3)$- group.

**Figure 2 | Structural relaxation behavior for glycols and water. a**. The *α*-relaxation times for the glycols measured to temperatures far above $T_g$. The symbols are the same as in Fig. 1. Dielectric relaxation data for water is also included as (▷)

[*37*] and (◁) [*38*]. **b.** Rescaled peak frequency data for the polyether glycol samples, where $Z=100 \cdot [-d\log(f_p)/d(1/T)]^{-1/2}$, $f_p$ is the α peak frequency from dielectric spectroscopy and $T^*$ [*14*] is a temperature characteristic of the observed dynamic crossover. The data and symbols are the same as in (a) and the solid lines are fits to the data in the regime near $T_g$ as described in the text. The numbers indicate the number of monomer units in the corresponding glycol chains, where $N=0$ corresponds to water. The solid line for water is the results of an extrapolation from the glycol data as described in c. The solid and dashed lines in region (I) are power law $f_p \propto (T-T_s)^\gamma$ fits to the water data shown in a with $T_s=217\pm1$ K and $\gamma=-2.04\pm0.05$. The solid line is the power law rescaled by $T^*=216$ K and the dashed line by 235 K. The dashed area thus marks the high-temperature behavior of water in this representation for the possible values of $T^*$. **c.** The negative slope $S$ (○) of the solid lines in region II and the value of the ratio $T^*/T_g$ (□) vs glycol molecular weight $M$. The dashed lines are exponential fits to the data. The solid lines are guides to the eye indicating the high $M$ behavior. The open black circle marks the extrapolated value of $S$ for water. The sections of the $T^*/T_g$ axis corresponding to the two suggested glass transition regimes for water are marked in blue, where the lower regime corresponds to a $T_g=165\pm5$ K and the upper regime to $124 \leq T_g \leq 136$ K.

**Figure 3 | $T_g$-behavior of polyethers.** $T_g$-values, defined as $T_g=T(\tau_\alpha=100$ s) for DME:s (○), MME:s (□) and glycols (△), respectively, plotted vs the number of molecular backbone atoms, n. We define n as all atoms, C and O, in the chain excluding C of the side-groups. The $T_g$-regime $124 \leq T_g \leq 136$ K for water, described further in the text, is indicated as a solid triangle with an errorbar. The DME with $N=0$, dimethyl ether, is a gas at room temperature and its $T_g$-value was estimated [*39*]

and the $T_g$-value for methanol was taken from literature [40]. The inset shows the same plot with the abscissa rescaled by the constant $\lambda$. The DME:s are plotted with $\lambda=1$ since they can not hydrogen bond to form extended structures. The MME data is plotted with $\lambda=2$, corresponding to paired oligomers and the glycol data with $\lambda=8$, thus assuming effective chains formed from on average 8 joined oligomers. The dashed lines are guides to the eye.

**Figure 4 | The *αβ*-relaxation scenario for supercooled water.** The dash-dotted line is a power law fit to the high temperature water data. The data and symbols are the same as in Fig. 2. The solid line indicates the VFT behavior estimated for water from our study of polyether glycols, as discussed in detail in the text. The open circles are structural relaxation data for water in a hydrogel matrix, where crystallization could be suppressed [4]. The triangles show relaxation data for confined water [27] and the dashed line is an Arrhenius function that fits the data, as further described in Fig. 5.

**Figure 5 | The secondary *β*-relaxation of polyethers and water. a.** $\beta$-relaxation time versus inverse temperature for the three polyether series, where red symbols denote DME:s, blue symbols MME:s and black symbols glycols with $N=1$ (○), $N=2$ (□), $N=3$ (△), $N=4$ (◇), $N=7$ (▷) and $N=70$ (▽). For $N=70$, the effects of chain-ends are negligible and all three series converge towards this behavior for long chains. The solid lines are Arrhenius fits to the different data sets. The dashed red line is an estimate of the DME monomer $\beta$-relaxation from a scaling analysis [13]. For the glycol monomer ($N=1$) the $\alpha$- and $\beta$-relaxations are merged. The result is a submerged $\beta$-relaxation observed only as a so called 'excess wing' on the high frequency flank of the structural relaxation [31]. The dash-dotted line shows an estimate of the position

of the submerged $\beta$-relaxation, see b, that is consistent with aging data where a separation of the two relaxations takes place [*31*]. Data for the main dielectric relaxation of water confined in a Na-vermiculate clay [*27*], where crystallinity is suppressed, is shown in triangles (◁). **b.** The relaxation time for the data sets in a**,** al**l** at a temperature of 143 K (marked with a dotted line in a), plotted vs molecular weight, *M*. The solid black circle marks the estimated data for the glycol monomer. The dotted line marks the time-scale for the glycol polymer for which the end-groups do not play a role. The dashed line is a power law, $\tau_\beta(T_i) \propto M^\gamma$ with $\gamma \cong -1.6$, fit to the *N*=2,3,4 and 7 glycol data.

**Figure 1**

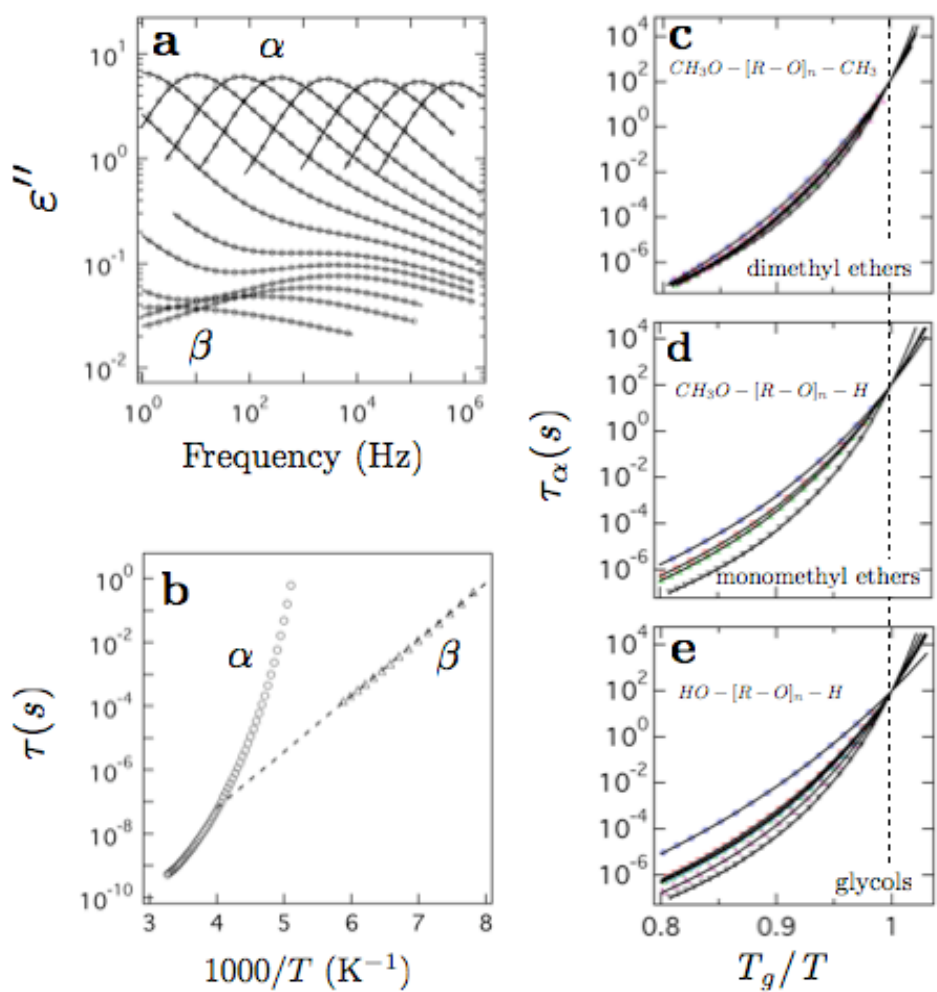

**Figure 2**

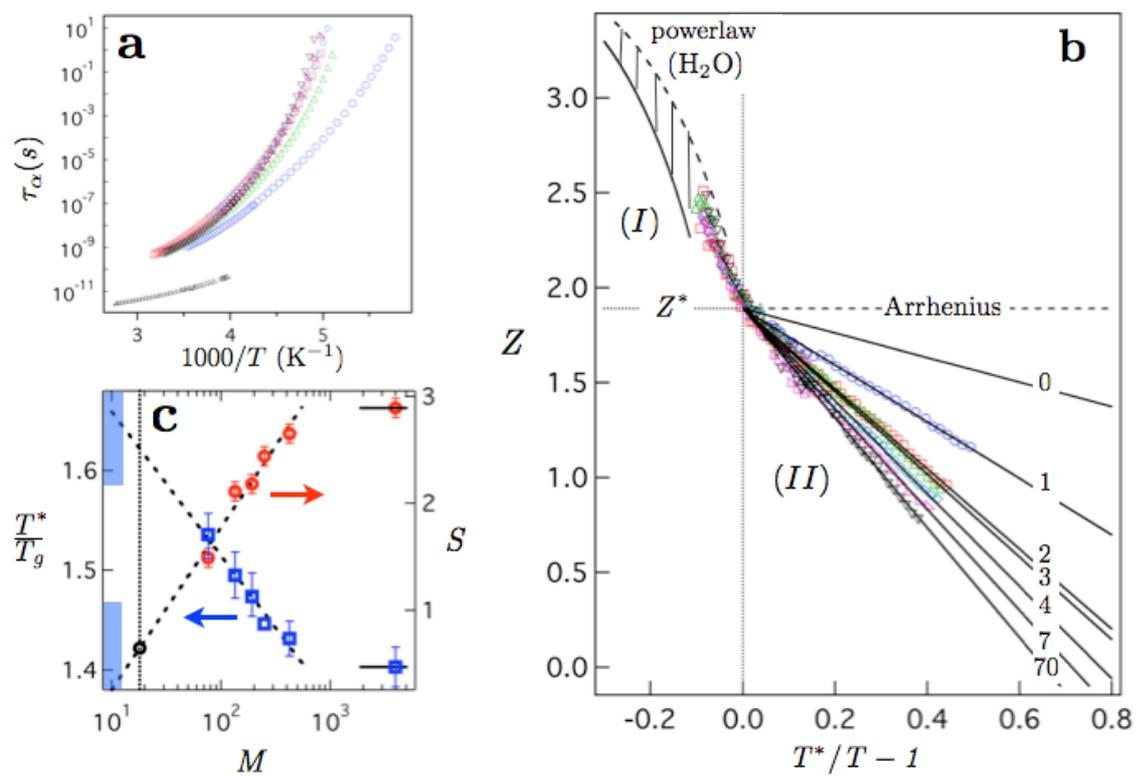

**Figure 3**

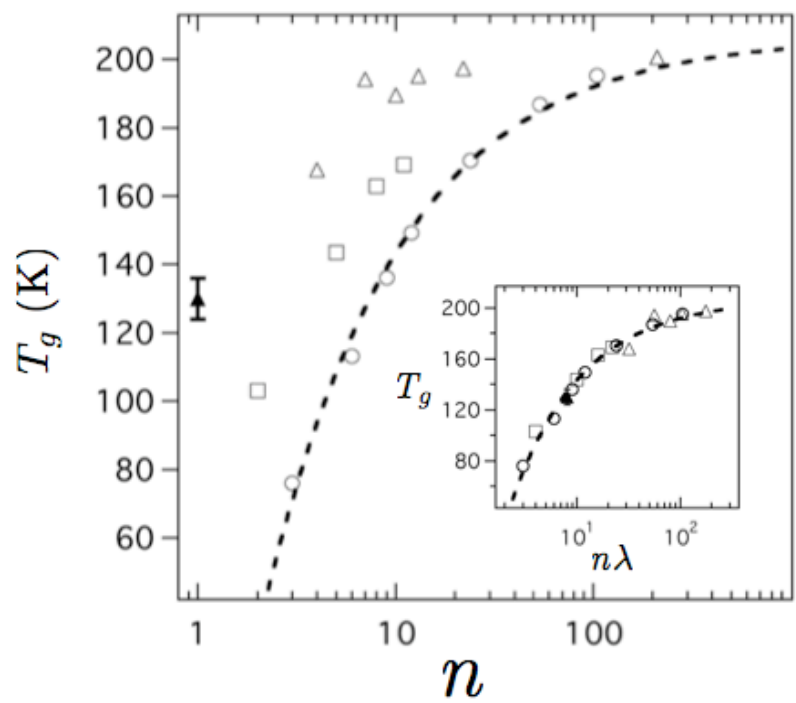

**Figure 4**

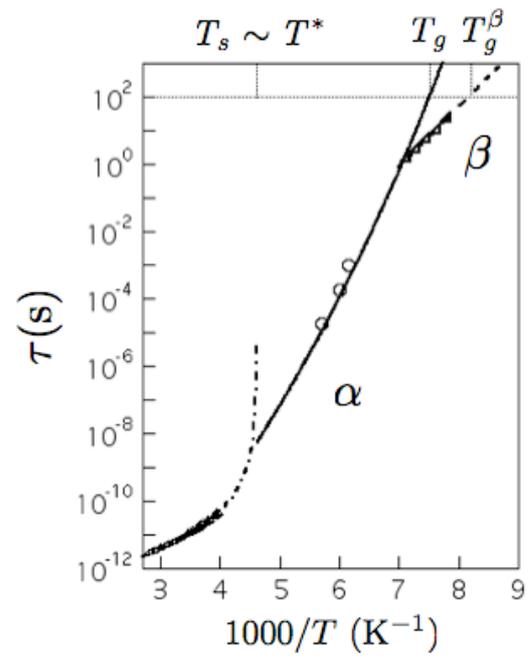

**Figure 5**

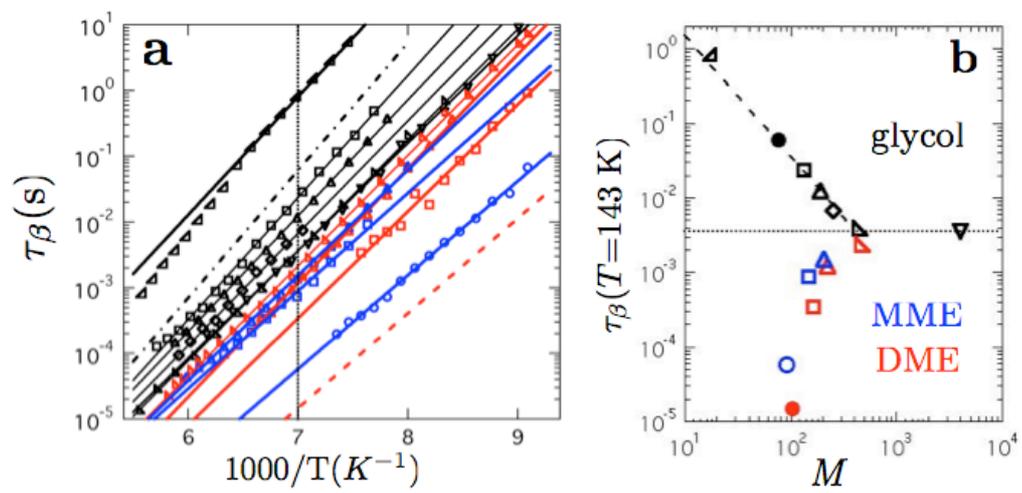